\documentclass[a4paper]{article}
\begin{document}
\topmargin -0.2cm \oddsidemargin -0.2cm \evensidemargin -1cm
\textheight 22cm \textwidth 12cm

\title{ Creation ion pairs and superconducting bosons.}
\author {Minasyan V.N.\\
Yerevan, Armenia}

\date{\today}

\maketitle
\begin{abstract}
First, it is shown that the creation of the spinless ion pairs in the lattice, which are hold by the binding with neighbor ion pairs together regarded as covalent. These ion pairs are created by the repulsive potential interaction of two ions which is bound as linear oscillator. The repulsive S-wave scattering between ion pairs 
and electrons is transformed to the attractive effective 
interaction between electrons which leads to a creation of electron pairs by 
a binding energy depending on the condensate fraction of ion 
pairs $\frac{N_0}{N}$. In this respect, the absence of 
ion pairs in the condensate destroys a binding energy 
of electron pairs and in turn so-called superconductimg phase. As new result presented theory is that the number of the superconducting bosons is not changed in the superconducting phase.
\end{abstract}


\vspace{130mm}

\vspace{5mm} 
 
{\bf  1. Introduction.} 
 
\vspace{5mm} 
Various models have been invoked in recent years [1-3], for
explaining macroscopic properties and the formation of high- $T_c$
superconducting phase. In 1928, Sommerfeld-Bloch had 
presented the model of individual electrons are been in the homogenous positive 
background of ions lattice, since the given system is  
electro-neutral. This model gives a good describing of the property  
metals, but it cannot explain a problem superconductivity. 

The charged ideal Bose gas was first investigated 
in 1955 by Schafroth [4], who defined a superconducting phase in 
the superconductor by using of a density of  bosons in the condensate. 

Following the initial work, connected with a charged nonideal 
Bose gas, proposed by Foldy [5], who worked at 
zero temperature ($T=0$) and calculated the energy and 
quasiparticle energy spectrum of the gas using a method presented by Bogoliubov [6].

To solve the problem of superconductivity, the Fr$\ddot o$lich had proposed to take into consideration the electron-lattice-electron attractive interaction [7], which proceeds when one electron interacts with the lattice and deforms it; a second electron takes this deformation. Thus, due to lattice deformation, there is appearance of an effective interaction between two electrons. In this respect, the ground state is separated from excited states by an energy gap near the Fermi level by theory of Bardeen, Cooper, Schrieffer and Bogoliubov (BCSB) [8] due to the creation so-called Cooper pairs [9]. The initial work is an experimental observation of the isotope effect [10] when the critical temperature $T_c$ of superconductors varies with the isotopic mass of ion $M$. The experimental date may be fitted by a relation of the form $ T_c=\frac{const}{\sqrt{M}}$. This fact implies that the role of lattice is a very important factor for solution of the problem superconductivity. 

In this letter, we propose a new approach for investigation the problem superconductivity, which differ from theories presented in above-mentioned references. In this respect, we show an every ion of lattice is catching on the neighbor ions with creating of the spinless ion pairs, which are as particles of a lattice chain.  On other hand, the interaction of electrons with ion pairs, due to a repulsive S-wave scattering which vanishes by inducing attractive interaction between two electrons.
As new result presented theory is that the number of 
superconducting bosons, which are a number of electron pairs 
in the condensate, is not changed in the superconducting phase. 

\vspace{5mm}
{\bf 2. Creation spinless ion pairs}.
\vspace{5mm}

The starting point our discussion is a presentation of the model superconductor as the stationary $n$ identical ions (which are a fermions) of lattice with charge $-e$, and mass $M$ in box of volume $V$, together with a background of $n$ electrons with mass $m_e$ and  opposite charge $e$ to preserve charge neutrality of system. The background of electrons represents as an electron gas. We suggest that the ions of lattice interact with each other by a repulsive potential energy of linear oscillator.

For beginning, we show that the electron + ion system is unstable in a superconductor. In analogy of the problem Hydrogen atom, the electron + ion system with opposite spins is bound as the one spinless pair with maximal binding energy:
 
$$
E_1=-\frac{m_e e^4}{4\hbar^2}
$$
where $m_e$ is the mass of electron.
The total energy of the electron + ion system is 
\begin{equation}
E=-\frac{m_e e^4}{2\hbar^2} +\frac{p^2}{2M}
\end{equation}

where $M$ is the mass of electron + ion system which represents as a neutral atom; $p$ is the momentum of  neutral atom. Obviously, at momenta $p\leq 2 p_f$ (where $p_f=\hbar \biggl(\frac{n}{V}\biggl)^{\frac{2}{3}}$ is the Fermi momentum of ion or electron), which is the maximal value of momentum for electrons and ions, at low temperatures, we obtain an important certainly condition for unstapled atom by support of (1), at  positive $E>0$:

$$
\frac{p_f^2}{2M}>\frac{m_e e^4}{2\hbar^2}
$$
which takes a following form:

$$
\frac{n}{V}\geq \biggl (\frac{m_e M e^4}{\hbar^4}\biggl)^{\frac{3}{2}}\approx 10^{22}sm^{-3}
$$

In this respect, we deal with metal because $\frac{n}{V}\geq 10^{22}sm^{-3}$ which can be considered as good superconductor.

Now, we consider the operator Hamiltonian $\hat{H}_s$ of two interacting neighboring ions of lattice within one-dimensional character:

\begin{equation}
\hat{H}_s= -\frac{\hbar^2}{2M}\Delta_{x_0}-
\frac{\hbar^2}{2M} \Delta_{x_1}+\hat{H}_{int}(x_0-x_1)
\end{equation}

where $x_0$ and $x_1$ are, respectively, the coordinates of two
 neighboring ions; 
$\hat{H}_{int}(x_0-x_1)$ is the repulsive potential energy of 
interaction between neighboring ions as a repulsive ion-ion scattering:

$$
\hat{H}_{int}(x)=\frac{K x^2}{2}
$$

where $ x=x_0-x_1$; $K$ is the rigidity modulus. 

The own wave function $\Psi_s (x_0,x_1)$ for 
the operator Hamiltonian $\hat{H}_s$ is introduced by a following way:
$$
\Psi_s(x_0,x_1)=e^
{i\frac{\vec{p}\vec{X}}{\hbar}}\psi_s(x)
$$
where $x=x_0-x_1$, $X=\frac{x_0 +x_1}{2}$ 

In this sense, the own quantity of the operator Hamiltonian $\hat{H}_s$ 
is the spectrum binding energy of the spinless pair of two ions with momentum $\vec{p}$: 

\begin{equation}
E_s=\frac{p^2}{2M_0}+ \varepsilon_s
\end{equation}

where $M_0=2 M$ is the mass of ion pair. 

The equation for $\varepsilon_s$ is presented as:

\begin{equation}
\biggl[-\frac{\hbar^2}{M}\Delta_{x}+M \omega^2_v x^2\biggl]
\psi_s(x)=\varepsilon_s \psi_s (x)
\end{equation}

where $\omega_v =\sqrt{\frac{K}{2M}}=\frac{const}{\sqrt{M}}$

We transform the form (4) in a following form:

\begin{equation}
\frac {d^2 \psi_s (x)}{dr^2} + \biggl (\lambda-\alpha^2 x^2 \biggl ) \psi_s(x)=0
\end{equation}

where we take $\alpha=-\frac{M\omega_v}{\hbar}$, and $\lambda =\frac{ M_0 \varepsilon_n }{\hbar^2}$

The (5) has a following solution by application of the wave function 
$\psi_s(x)$ trough Chebishev-Hermit function 
$H_s (it)$ from an imaginary number as argument  $it$ [11] (where 
$i$ is the imaginary one; $t$ is the real number; $s=0;1;2;\cdots$) as:

$$
\psi_s(\vec{r})=e^{-\alpha\cdot x^2}H_s(\sqrt{\alpha }\cdot x) 
$$
where we take $\alpha=-\frac{M\omega_v}{\hbar}$; 
$$
H_s(it)=i^s e^{-t^2}\frac{d^s e^{t^2}}{d t^s}
$$ 

and

$$
\lambda=\alpha (s+\frac{1}{2})
$$

Consequently, the quantity of the binding energy of ion pair $\varepsilon_s$ presents: 
\begin{equation}
\varepsilon_s =-\hbar\omega_v\biggl(s+\frac{1}{2}\biggl)
\end{equation}

Inserting (6) into (3) leads to finding of the total energy of system describing of moving ion pair:

\begin{equation}
E_s =-\hbar\omega_v\biggl(s+\frac{1}{2}\biggl) +\frac{p^2}{2 M_0} 
\end{equation}
 
The normal state of ion pair corresponds to quantity  $s=0$ which defines the maximal quantity of the binding energy of ion pair: 
\begin{equation}
E_0 =-\frac{\hbar\omega_v}{2}+\frac{p^2}{2 M_0}
\end{equation}

 As we mentioned the single ion represent as fermion, therefore, the momentum pair of ion $p$ has to satisfy to the condition $p\leq 2p_f$, at low temperatures. In this context, if all ions of lattice are identical, then the stability of spinless ion pair is determined by condition $E_0=0$ in (8) which leads to a following expression:

\begin{equation}
M_0\hbar\omega_v=\hbar^2\biggl(\frac{8\pi N}{3V}\biggl)^{\frac{2}{3}}
\end{equation}
where $2M\hbar\omega_v =const \sqrt {M}$, at $M_0=2M$, $N=\frac{n}{2}$

Thus, we find an important expression, which is need 
for explaining the isotope effect
\begin{equation}
\biggl(\frac{N}{V}\biggl)^{\frac{2}{3}}=const\sqrt{M}
\end{equation}

Hence, we can state that every ion of lattice is bound with neighbor ion regarded as covalent. 

Thus, the gas of ions of lattice is transformed to the charged 
spinless Bose gas consisting of $N=\frac{n}{2}$ identical pairs with 
charge $e_i=-2e$ and mass $ M_0=2M$ which represent as particles of lattice chain because they can be moved only by lines of lattice.   

In this respect, the Hamiltonian of a charged ideal Bose gas consisting of ion pairs is expressed in the second quantization form as [3]:

\begin{equation}
\hat{H}_i= \sum_{\vec{p}}\frac{p^2}{2M_0}
\hat{a}^{+}_{\vec{p}}\hat{a}_{\vec{p}}
\end{equation}

where $\hat{a}^{+}_{\vec{p}}$ and $\hat{a}_{\vec{p}}$ are the
"creation" and  "annihilation" operators of an ion pairs with 
momentum $\vec{p}$.

\vspace{5mm}
{\bf 3. The effective attractive potential interaction between two electrons. }
\vspace{5mm}

Now, consider a ion pair gas-electron gas mixture as a Bose gas of $N$ identical spinless charged ion pairs with mass $M_0$ and charge $e_i$ in box of volume 
$V$. These ion pairs in the chain interact with a Fermi gas of $n$ electrons by a repulsive S-wave scattering. To describe an ion pair gas-electron gas mixture we apply the well-known Fr$\ddot o$lich approach [13]. At solving of the problem superconductivity, the Fr$\ddot o$lich described a electron gas-phonon gas mixture where exist an interaction between electrons (fermions) and phonons (bosons) of lattice. Due to an application of the Fr$\ddot o$lich method we arrive to the Hamiltonian of an ion pair gas-electron gas mixture which is expressed in the second quantization form:

\begin{equation}
\hat{H}= \sum_{\vec{p}}\frac{p^2}{2M_0}
\hat{a}^{+}_{\vec{p}}\hat{a}_{\vec{p}}+
\frac{1}{2V}\sum_{\vec{p}\not=0}U_{\vec{p}}\hat{\varrho}_{\vec{p},i}
\hat{\varrho}_{-\vec{p},e}+\sum_{\vec{p},\sigma }\frac{p^2}{2m_e}
\hat{b}^{+}_{\vec{p},\sigma }\hat{b}_{\vec{p},\sigma }
\end{equation}

where $U_{\vec{p}}$ is the Fourier
transform of the attractive S-wave scattering 

\begin{equation}
U_{\vec{p}} = -\frac{4\pi d\hbar^2}{\mu}= U_0 
\end{equation}
where $d$ is the scattering amplitude; $\mu =\frac{M_0 m_e}{ M_0 +m_e }\approx m_e$ because $M_0\gg m_e$ 

$\hat{b}^{+}_{\vec{p},\sigma}$ and 
$\hat{b}_{\vec{p},\sigma }$ are, respectively,  the operators of creation and 
annihilation for free fermion with momentum $\vec{p}$, by the value of its 
spin z-component $\sigma=^{+}_{-}\frac{1}{2}$. These operators satisfy to 
the Fermi commutation relations $[\cdot\cdot\cdot]_{+}$ as:

\begin{equation}
\biggl[\hat{b}_{\vec{p},\sigma}, \hat{b}^{+}_{\vec{p}^{'},
\sigma^{'}}\biggl]_{+} =
\delta_{\vec{p},\vec{p^{'}}}\cdot\delta_{\sigma,\sigma^{'}}
\end{equation}

\begin{equation}
[\hat{b}_{\vec{p},\sigma}, \hat{b}_{\vec{p^{'}}, \sigma^{'}}]_{+}= 0
\end{equation}

\begin{equation}
[\hat{b}^{+}_{\vec{p},\sigma}, \hat{b}^{+}_{\vec{p^{'}}, 
\sigma^{'}}]_{+}= 0
\end{equation}

The density operator of electrons with spin $\sigma$ in momentum 
$\vec{p}$ is defined as
\begin{equation}
\hat{\varrho}_{\vec{p},e}=\sum_{\vec{p}_1, \sigma }
\hat{b}^{+}_{\vec{p}_1-\vec{p},\sigma }\hat{b}_{\vec{p}_1,\sigma }
\end{equation}
where $\hat{\varrho}^{+}_{\vec{p},e}=\hat{\varrho}_{-\vec{p},e}$

The operator of total number of electrons is
\begin{equation}
\sum_{\vec{p},\sigma}\hat{b}^{+}_{\vec{p},\sigma}\hat{b}_{\vec{p},\sigma}=\hat{n}
\end{equation}

In this context, the density operator of ion pairs is presented by following form: 

\begin{equation}
\hat{\varrho}_{\vec{p},i}=\sum_{\vec{p}_1}
\hat{a}^{+}_{\vec{p}_1-\vec{p}}\hat{a}_{\vec{p}_1}
\end{equation}

where the Bose commutation relations are presented as:

\begin{equation}
[\hat{a}_{\vec{p}^{'}},\hat{a}^{+}_{\vec{p}^{"}}]=\delta_{\vec{p}^{'},\vec{p}^{"}}
\end{equation}

\begin{equation}
[\hat{a}_{\vec{p}^{'}},\hat{a}_{\vec{p}^{"}}]=0
\end{equation}

\begin{equation}
[\hat{a}^{+}_{\vec{p}^{'}},\hat{a}^{+}_{\vec{p}^{"}}]=0
\end{equation}

We note that the operator for bosons $\hat{a}_{\vec{p}}$  and the one for fermion 
 $\hat{b}_{\vec{p},\sigma }$ is an independent.

Now, according to the Bogoliubov's theory [6], the operators $\hat{a}_0$ and  
$\hat{a}^{+}_0$ are replaced by c-numbers  $\hat{a}_0=\hat{a}^{+}_0=\sqrt{N_0}$  
within approximation of a macroscopic number of  ion pairs in the condensate $N_0\gg 1$ we introduced the definition of a Bose 
condensation [12]  proposed by the Penrose-Onsager  

\begin{equation}
\lim_{N_0, N\rightarrow\infty}\frac{N_0}{N}=const
\end{equation} 

The equation (23) allows us to introduce an important assumption for an occupation number $ N_{\vec{p}}$ of bosons with momentum  $\vec{p}$, which are $ N_{\vec{p}\not 0}<<N_0$: 

\begin{equation}
\lim_{N_0\rightarrow\infty}\frac{N_{\vec{p}}}{N_0}=\delta_{\vec{p},0}
\end{equation}

which defines a property of operators 
$\frac{\hat{a}^{+}_{\vec{p}_1-\vec{p}}}{\sqrt{N_0}}$, 
$\frac{\hat{a}_{\vec{p}_1-\vec{p}}}{\sqrt{N_0}}$ by applying (20). Obviously,

\begin{equation}
\lim_{N_0\rightarrow\infty}\frac{\hat{a}^{+}_{\vec{p}_1-\vec{p}}}{\sqrt{N_0}}=\delta_{\vec{p}_1, \vec{p}} 
\end{equation} 
and
\begin{equation}
\lim_{N_0\rightarrow\infty}\frac{\hat{a}_{\vec{p}_1-\vec{p}}}{\sqrt{N_0}}=
\delta_{\vec{p}_1, \vec{p}} 
\end{equation}

Thus, we reach to the density operators of bosons $\hat{\varrho}_{\vec{p}}$ and 
$\hat{\varrho}^{+}_{\vec{p}}$ presented by Bogoliubov [6] 
without using of a definition of Bose condensation as $N_0\approx N$:

\begin{equation}
\hat{\varrho}_{\vec{p},i }= \sqrt{N_0}\biggl(\hat{a}^{+}_{-\vec{p}}+
\hat{a}_{\vec{p}}\biggl)
\end{equation}

and
\begin{equation}
\hat{\varrho}^{+}_{\vec{p},i }=\sqrt{N_0}\biggl(\hat{a}_{-\vec{p}}+
\hat{a}^{+}_{\vec{p}}\biggl)
\end{equation}
which works on the condition $\frac{N_0}{N}=const$

Thus, an application of (27) and (28) into (12) reduces the Hamiltonian of system $\hat{H}$ by a following form:

\begin{equation}
\hat{H}= \sum_{\vec{p}}\frac{p^2}{2M_0}
\hat{a}^{+}_{\vec{p}}\hat{a}_{\vec{p}}+
\frac{1}{2V}\sum_{\vec{p}}U_0\sqrt{N_0}\biggl(\hat{a}^{+}_{-\vec{p}}+\hat{a}_{\vec{p}}\biggl)\hat{\varrho}_{-\vec{p},e}+\sum_{\vec{p},\sigma }\frac{p^2}{2m_e }\hat{b}^{+}_{\vec{p},\sigma }\hat{b}_{\vec{p},\sigma }
\end{equation}

To allocate anomalous terms in the Hamiltonian of 
system in (29), we apply the Fr$\ddot o$lich approach [13] which allows to do the canonical transformation for
an operator $\hat{H}_n$ within introducing a new operator $\tilde{H}$:

\begin{equation}
\tilde{H}=\exp\biggl(\hat{S}^{+}\biggl)\hat{H}
\exp\biggl(\hat{S}\biggl)
\end{equation}

which can presents as:

\begin{equation}
\tilde{H}=\exp\biggl(\hat{S}^{+}\biggl)\hat{H}
\exp\biggl(\hat{S}\biggl) = \hat{H}-[\hat{S},\hat{H}]+
\frac{1}{2}[\hat{S},[\hat{S},\hat{H}]]-\cdots
\end{equation}

where 
$$
\hat{S}^{+}=\sum_{\vec{p}}\hat{S^{+}_{\vec{p}}}
$$
and   
$$
\hat{S}=\sum_{\vec{p}}\hat{S_{\vec{p}}}
$$
 with condition $\hat{S}^{+} = -\hat{S}$

We assume that

\begin{equation}
\hat{S_{\vec{p}}}=A_{\vec{p}}\biggl (\hat{\varrho}_{\vec{p},e}
\hat{a}_{\vec{p}}-
\hat{\varrho}^{+}_{\vec{p},e}\hat{a}^{+}_{\vec{p}}\biggl)
\end{equation}

where $A_{\vec{p}}$ is the unknown  real symmetrical function from  a momentum
$\vec{p}$ which we will find. In this respect, substituting (29) and (32) into  (31), and making a some transformations, we find a following terms involving into (31):

\begin{equation}
[\hat{S},\hat{H}]=-\frac{1}{V}\sum_{\vec{p}} A_{\vec{p}}U_0\sqrt{N_0}\hat{\varrho}_{\vec{p},e}\hat{\varrho}_{-\vec{p},e}-\sum_{\vec{p}} A_{\vec{p}} \frac{p^2}{2M_0}\hat{\varrho}_{\vec{p},e}\biggl(\hat{a}^{+}_{-\vec{p}}+\hat{a}_{\vec{p}}\biggl)
\end{equation}

\begin{equation}
\frac{1}{2}[\hat{S},[\hat{S},\hat{H}]]=-\sum_{\vec{p}} A^2_{\vec{p}} \frac{p^2}{2M_0}\varrho_{\vec{p},e}\hat{\varrho}_{-\vec{p},e}
\end{equation}

and $[\hat{S}, [\hat{S},[\hat{S},\hat{H}]]]=0$ within a Bose commutation relation 
$[\varrho_{\vec{p}_1,e},\hat{\varrho}_{\vec{p}_2,e}]=0$; $[\hat{b}^{+}_{\vec{p}_1,\sigma}\hat{b}_{\vec{p}_1,\sigma},\hat{\varrho}_{\vec{p}_2,e}]=0$. 
Then, the form of new operator $\tilde{H}$ in (31) is reduced to a following form:

\begin{eqnarray}
\tilde{H}& =&\sum_{\vec{p}}\frac{p^2}{2M_0}
\hat{a}^{+}_{\vec{p}}\hat{a}_{\vec{p}}+
\frac{1}{2V}\sum_{\vec{p}}U_0\sqrt{N_0}\biggl(\hat{a}^{+}_{-\vec{p}}+\hat{a}_{\vec{p}}\biggl)\hat{\varrho}_{-\vec{p},e}+
\nonumber\\
&+&\sum_{\vec{p},\sigma }\frac{p^2}{2m_e}\hat{b}^{+}_{\vec{p},\sigma }\hat{b}_{\vec{p},\sigma }+
\frac{1}{V}\sum_{\vec{p}} A_{\vec{p}}U_0\sqrt{N_0}\hat{\varrho}_{\vec{p},e}\hat{\varrho}_{-\vec{p},e}+
\nonumber\\
&+&\sum_{\vec{p}} A_{\vec{p}} \frac{p^2}{2M_0}\biggl(\hat{a}^{+}_{-\vec{p}}+\hat{a}_{\vec{p}}\biggl) \hat{\varrho}_{-\vec{p},e}-\sum_{\vec{p}} A^2_{\vec{p}} \frac{p^2}{2M_0}\hat{\varrho}_{\vec{p},e}\hat{\varrho}_{-\vec{p},e}
\end{eqnarray}

Removing of the terms of second and fifth in right part of (35), we obtain a quantity for $A_{\vec{p}}$:

\begin{equation}
A_{\vec{p}}=-\frac{U_0 M_0 \sqrt{N_0}}{p^2 V}
\end{equation}

which within introducing it to a fourth and sixth terms in right part of (35) leads to finding the form of new operator $\tilde{H}$ without involving of an   
Interaction part between pairs of ion and electrons:

\begin{equation}
\tilde{H}= \sum_{\vec{p}}\frac{p^2}{2M_0}
\hat{a}^{+}_{\vec{p}}\hat{a}_{\vec{p}}+\sum_{\vec{p},\sigma }\frac{p^2}{2m_e }\hat{b}^{+}_{\vec{p},\sigma }\hat{b}_{\vec{p},\sigma }+
\frac{1}{2V}\sum_{\vec{p}}V_{\vec{p}}
\hat{\varrho}_{\vec{p},e}\hat{\varrho}_{-\vec{p},e}
\end{equation}

where $V_{\vec{p}}$ is the effective attractive potential of interaction between electrons in momemtum space: 

$$
V_{\vec{p}}=-\frac{3 M_0 U^2_0 N_0}{V p^2}
$$

which presents by the term of an effective charge $ e_*$ as :

\begin{equation}
V_{\vec{p}}=-\frac{4\pi \hbar^2 e^2_*}{p^2}
\end{equation}

where

$$
e_*=\frac{U_0}{2\hbar }\sqrt{\frac{3 M_0 N_0}{V \pi }}
$$

The transformation of an effective attractive potential between electrons $V_{\vec{p}}$ in the momentum space in (38) to the one of a coordinate space is presented by following formulae:

\begin{equation}
V (\vec{r})=\frac{1}{V}\sum_{\vec{p}} V_{\vec{p}}\cdot e^{i\frac{\vec{p}\vec{r}}{\hbar}}=-\frac{e^2_*}{r}
\end{equation}

As we see, there is an appearance of an effective attractive potential of interaction between two electrons in the coordinate space $V (\vec{r})$.

Now, consider an interacting electron gas of $n$ identical electrons with mass $m_e$ and charge $e$ which interact with each other by an attractive potential $V (\vec{r})=-\frac{e^2_*}{r}$. In analogy of the problem Hydrogen, two electrons with opposite spins is bound as one spinless pair with binding energy:
 
\begin{equation}
E_l=-\frac{m_e e^4_*}{4\hbar^2 l^2}=-const \frac{N^2_0}{N^2}
\end{equation}

where $l$ is the natural number; $ const >0$

Thus, there is a charged Bose gas consisting of $N=\frac{n}{2}$ spinless electron pairs with mass $m=2m_e$ and charge $e_0=2e$.

In this respect, the Hamiltonian of an ideal Bose gas, consisting of electron pairs, is expressed in the second quantization form as [3]:

\begin{equation}
\hat{H}_e= \sum_{\vec{p},\sigma }\frac{p^2}{2m_e }\hat{b}^{+}_{\vec{p},\sigma }\hat{b}_{\vec{p},\sigma }+
\frac{1}{2V}\sum_{\vec{p}}V_{\vec{p}}
\hat{\varrho}_{\vec{p},e}\hat{\varrho}_{-\vec{p},e}=\sum_{\vec{p}}\frac{p^2}{2m}
\hat{d}^{+}_{\vec{p}}\hat{d}_{\vec{p}}
\end{equation}

where $\hat{d}^{+}_{\vec{p}}$ and $\hat{d}_{\vec{p}}$ are the
"creation" and  "annihilation" operators of a free electron pairs  with 
momentum $\vec{p}$.

Then, the Hamiltonian of system in (37) can write in a following form:

\begin{equation}
\hat{H}= \sum_{\vec{p}}\frac{p^2}{2M_0}
\hat{a}^{+}_{\vec{p}}\hat{a}_{\vec{p}}+\sum_{\vec{p}}\frac{p^2}{2m}
\hat{d}^{+}_{\vec{p}}\hat{d}_{\vec{p}}
\end{equation}

\vspace{5mm}
{\bf 3. Results and Discussion.}
\vspace{5mm}

We proceed the investigation of a thermodynamic property of an ideal ion and electron pairs presented independent gases. 
In statistical equilibrium, the equations for the densities of ion and electron pairs in 
the condensate  are represented as: 

\begin{equation}
\frac{N_{0}}{V}=\frac{N}{V}-\frac{1}{V}\sum_{\vec{p}}
\overline{\hat{a}^{+}_{\vec{p}}\hat{a}_{\vec{p}}}
\end{equation}

\begin{equation}
\frac{N_{0,e}}{V}=\frac{N}{V}-\frac{1}{V}\sum_{\vec{p}}
\overline{\hat{d}^{+}_{\vec{p}}\hat{d}_{\vec{p}}}
\end{equation}

where $\overline{\hat{a}^{+}_{\vec{p}}\hat{a}_{\vec{p}}}$ and $\overline{\hat{d}^{+}_{\vec{p}}\hat{d}_{\vec{p}}}$, receptively,
are the averages numbers of ion and electron pairs with the momentum $\vec{p}$:

\begin{equation}
\overline{\hat{a}^{+}_{\vec{p}}\hat{a}_{\vec{p}}}=
\frac{1}{e^{\frac{p^2 }{2M_0 kT}}-1}
\end{equation}

and
\begin{equation}
\overline{\hat{d}^{+}_{\vec{p}}\hat{d}_{\vec{p}}}=
\frac{1}{e^{\frac{p^2 }{2m kT}}-1}
\end{equation}

Consequently, inserting (45) into (43), and (46) into (44), we reach to an important expressions:  

\begin{equation}
\frac{N_{0}}{N}=1-\biggl (\frac{T}{T_{c}}\biggl)^{\frac{3}{2}}
\end{equation}
and

\begin{equation}
\frac{N_{0,e}}{N}=1-\biggl (\frac{T}{T_{c,e}}\biggl)^{\frac{3}{2}}
\end{equation}

where $T_{c}$ and $T_{c,e}$, respectively, are the transition temperatures for an ion  and electron pairs of an ideal Bose gases, which are presented as:

\begin{equation}
T_{c}=\frac{1.6\hbar^2}{M k}\biggl(\frac{N}{V}\biggl)^{\frac{2}{3}}=
\frac{const}{\sqrt{M}}
\end{equation}
and

\begin{equation}
T_{c,e}=\frac{1.6\hbar^2}{m_e k}\biggl(\frac{N}{V}\biggl)^{\frac{2}{3}}
\end{equation}
As result of (49) and (50) is $ T_{c,e} \gg T_{c}$ because $ M \gg m_e$. This fact implies that a condensation of ion pairs is absence a quick than it could be occur with an electron pairs. Therefore, the equation (48) is fulfilled by temperatures varying in state $0\leq T\leq T_{c}$ with condition $\frac{ T_{c}}{ T_{c,e}}=\frac{ m_e}{ M}\ll 1$. In this respect, in the state of superconducting phase, all electron pairs are been in the condensate $N_{0,e}\approx N$, as result of (48), therefore, we can call an electron pairs in the condensate as a superconducting bosons. Also, we can see that the critical temperature $ T_c$ of superconductors is defined by an absence the fraction of ion pairs in the condensate $\frac{ N_0}{ N}=0$ which destroys the binding energy of electron pair $E_l=0$ in (40).    

Obviously, the application of equation (10) into (49) leads to explaining of an isotope effect because it confirms an existence of ion pairs. At least, the obtained result for quantity $T_{c}$ is satisfied at least quantitatively 
by other experimental dates because at $\frac{N}{V}\approx 10^{28}m^{-3}$ 
for many metal, $T_{c}\approx 1K$. 

\vspace{5mm}
{\bf 5. Conclusion.}
\vspace{5mm}

The problem of superconductivity is a very complex, and as yet not fully solved, 
problem. Apart from the intrinsic value of solving a particular many -body problem, 
a solution is desirable as a creation pairs of ion, which can 
serve as a model for a superconductors. In this context, we proved that the electron gas interacting with phonons of lattice couldn't represent as a model for superconductors that were proposed by BCSB because a repulsive potential interaction, having one-dimensional character between ions, creates of ion pair. Further, the interaction between ion pairs and electrons induces electron pairs. The evidence of existence ion pairs is confirmed yet by the isotope effect presented in (49). The absence of ion pairs in the condensate destroys a superconductimg phase in the superconductors. As new result presented theory is that the number of the superconducting bosons is not changed in the superconducting phase

\newpage

\begin{center}
{\bf References}
\end{center}
\begin{enumerate}
\item
P.W.~Anderson~., The theory of superconductivity in the high
-$T_c$ cuprates (Prinston University Press,Princeton, New Jersey,
1997).
\item
A.S.~Alexandrov~and Sir Nevil ~Mott~,"Polarons  and Bipolarons",
Cambridge- Loughborough, Word Scientific, (1995).
\item
Gerald D.~Mahan~, "Many-Particle Physics", ~Plenium~ Press New  York,  (1990); Kerson ~Huang~, "Statistical Mechanics", ~John Willey, New
York-London, (1963)
\item
M.R.~Schafroth~, Phys.Rev., {\bf 100}, 463 (1955).
\item
L.L.~Foldy~, Phys.Rev., {\bf 124}, 649 (1961).
\item
N.Bogoliubov, Jour. of Phys.(USSR),  11, 23 (1947)
\item
H.~Fr$\ddot o$lich~, Phys.Rev., {\bf 79},  845, (1950); 
Proc.Roy. Soc, {\bf 215}, 291 (1952).
\item
J.~Bardeen~, L.N.~Cooper~, and J.R.~Schrieffer~, Phys.Rev.{\bf
108},~1175~(1957); N.N.~Bogoliubov~, ~Nuovo~ ~Cimento~, ~{\bf 7},~794~(1958).
\item
L.N.~Cooper~, Phys.Rev., {\bf 104}, 1189 (1956).
\item
~Garland J.W., Jr.,~, Phys.Rev.Lett., {\bf 11}, 114 (1963); W.I.~McMillan.,~ Phys.Rev., {\bf 167}, 331 (1968)
\item 
M.A.~Lavrentiev~, and B.V. ~Shabat ~ "Nauka", ~ Moscow,~, (1973)
\item 
O. ~Penrose~ and L. ~Onsager~, ~Phys. Rev.,~{\bf 104}~, ~576~ (1956)
\item
H.~Fr$\ddot o$lich~, Phys.Rev., {\bf 79},  845, (1950); 
Proc.Roy. Soc, {\bf 215}, 291 (1952).

\end{enumerate} 
\end{document}